\documentclass[12pt]{article}

\begin{document}
\title{{\bf QCD Sum Rules: Intercrossed Relations for the
$\Sigma^{0}-\Lambda$ Mass Splitting
}}
\author{
{ A.\"{O}zpineci}\thanks{ozpineci@ictp.trieste.it} \\
{\small International Centre for Theoretical Physics,} \\ 
{\small Strada Costiera 11,I34100, Trieste, Italy} \\
\and
{ S.B.~Yakovlev, and V.S.~Zamiralov} \thanks{zamir@depni.sinp.msu.ru} \\
{\small Institute of Nuclear Physics,} \\
{\small M.V. Lomonosov Moscow State University,} \\
{\small Vorobjovy Gory, Moscow, Russia.}
}
\begin{titlepage}
\maketitle
\thispagestyle{empty}
\begin{abstract}
New relations between QCD Borel sum rules for masses of
$\Sigma^{0}$ and $\Lambda$ hyperons are constructed.
It is shown that starting from the sum rule for the
$\Sigma^{0}$ hyperon mass it is straightforward to obtain
the corresponding sum rule for the $\Lambda$ hyperon mass
and {\it vice versa}.
\\
\\
{\em PACS:} 11.30.Hv, 11.55.Hx \\
{\em Keywords}: Octet Baryons, Mass sum rules, $SU(3)_f$ Symmetry 
\end{abstract}
\end{titlepage}


\section{Introduction}
Recently a series of papers were dedicated to study hadron 
properties of the $\Sigma, \Sigma_{c}$ baryons
as well as of the $\Lambda, \Lambda_{c}$ ones in the framework
of various QCD sum rules \cite{Hwang,Hwang2,Altug1,Altug,Hwang1,lee}
which have their 
origin in the works \cite{Ioffe,Ioffe2,Yung}. In \cite{Ioffe,Yung}, the nucleons were studied using the
QCD sum rules approach and in \cite{Pasu,Pasu2}, the study was extended to the whole baryon octet. In \cite{Ioffe2}, the whole baryon octet was studied using QCD sum rules together 
with Gell-Mann-Okubo relation to obtain the mass of the $\Lambda$.
Many interesting results were obtained. But full expressions for
mass or magnetic moment sum rules often become too long and tedious to 
achieve and prove. Is it possible to relate all these results 
among themselves and
derive, say, $\Lambda$ hyperon properties from that of $\Sigma$ ones
and {\it vice versa} or just to check them mutually?


We propose here  nonlinear intercrossed relations which relate
matrix elements of $\Sigma$-like baryons with those of
$\Lambda$-like ones  and {\it vice versa}. Their origin lies in
the relation between isotopic, $U$- and $V$-spin quantities
and is quasi obvious in the framework of the quark model.
These relations prove to be valid for any QCD sum rules
and seem to be useful while obtaining hadron properties
of the $\Lambda$-like baryons from those of the
$\Sigma$-like baryons ( and {\it vice versa}) or checking
expressions for them reciprocally. 
The latter proves to be important as final QCD SR's comes to
be rather long and cumbersome so
it becomes a difficult work to compare or prove them term by term.
\section{Relation between magnetic moments of
hyperons $\Sigma^{0}$  and $\Lambda$ in the NRQM}

We begin with a simple example. Let us write magnetic moments
of hyperons $\Sigma^{0}$  and $\Lambda$ of the baryon octet
in the NRQM:
\begin{equation}
\mu(\Sigma^{0}(ud,s))=\frac{2}{3}\mu_{u}+\frac{2}{3}\mu_{d}-
\frac{1}{3}\mu_{s};\quad \mu(\Lambda)=\mu_{s}.
\label{mm8}
\end{equation}
As it is known magnetic moment of any other baryon of the octet but that of
the $\Lambda$ hyperon can be obtained from the expression
for the  $\Sigma^{0}$. E.g.,  magnetic moment of the  $\Sigma^{+}(uu,s)$
hyperon is obtained just by putting $\mu_{u}$ instead of $\mu_{d}$
in Eq.(\ref{mm8}):
$$
\mu(\Sigma^{+})=\frac{4}{3}\mu_{u}-\frac{1}{3}\mu_{s}.
$$
But magnetic moment of the $\Lambda$ hyperon can be also obtained
from that of the $\Sigma^{0}$ one, as well as 
 magnetic moment of the  $\Sigma^{0}$  can be obtained 
from that of the $\Lambda$ one. For that purpose
let us formally perform in Eq.(\ref{mm8}) the exchange $d\leftrightarrow s$ 
to get
\begin{equation}
\mu(\tilde\Sigma^{0}_{d\leftrightarrow s})=
\frac{2}{3}\mu_{u}+\frac{2}{3}\mu_{s}-
\frac{1}{3}\mu_{d};\quad \mu(\tilde\Lambda_{d\leftrightarrow s})=\mu_{d}
\label{sigma}
\end{equation}
and the exchange $u\leftrightarrow s$ to get
\begin{equation}
\mu(\tilde\Sigma^{0}_{u\leftrightarrow s})=
\frac{2}{3}\mu_{d}+\frac{2}{3}\mu_{s}-
\frac{1}{3}\mu_{u};\quad \mu(\tilde\Lambda_{u\leftrightarrow s})=\mu_{u}.
\label{lambda}
\end{equation}
The following relations are valid:
\begin{eqnarray}
\label{nrqmrelations}
2(\mu(\tilde\Sigma^{0}_{d\leftrightarrow s})+
\mu(\tilde\Sigma^{0}_{u\leftrightarrow s}))-
\mu(\Sigma^{0})=3\mu(\Lambda);\\
2(\mu(\tilde\Lambda_{d\leftrightarrow s})+
\mu(\tilde\Lambda_{u\leftrightarrow s}))-
\mu(\Lambda)=3\mu(\Sigma^{0})\nonumber.
\end{eqnarray}
The origin of these relations lies in the structure of
baryon wave functions in the NRQM with isospin $I=1,0$
and $I_{3}=0$:
$$
2\sqrt{3}|\Sigma^{0}(ud,s)>_{\uparrow}=
$$
$$
|2u_{\uparrow}d_{\uparrow}s_{\downarrow}+
2d_{\uparrow}u_{\uparrow}s_{\downarrow}-
u_{\uparrow}s_{\uparrow}d_{\downarrow}-
s_{\uparrow}u_{\uparrow}d_{\downarrow}-
d_{\uparrow}s_{\uparrow}u_{\downarrow}-
s_{\uparrow}d_{\uparrow}u_{\downarrow}>,
$$
$$
2|\Lambda>_{\uparrow}=
|d_{\uparrow}s_{\uparrow}u_{\downarrow}+
s_{\uparrow}d_{\uparrow}u_{\downarrow}-
u_{\uparrow}s_{\uparrow}d_{\downarrow}-
s_{\uparrow}u_{\uparrow}d_{\downarrow}>,
$$
where $q_{\uparrow}$ ($q_{\downarrow}$) means wave function of
the quark $q$ (here $q=u,d,s$) with the helicity +1/2 (-1/2).
With the exchanges $d\leftrightarrow s$ and $u\leftrightarrow s$ 
one arrives at the corresponding $U$-spin and $V$-spin quantities, so
$U=1,0$ and $U_{3}=0$ baryon wave functions are
$$
-2|\tilde\Sigma^{0}_{d\leftrightarrow s}(us,d)>=
|\Sigma^{0}(ud,s)>+\sqrt{3}|\Lambda>,
$$
$$
-2|\tilde\Lambda_{d\leftrightarrow s}>=
-\sqrt{3}|\Sigma^{0}(ud,s)>+|\Lambda>,
$$
while $V=1,V_{3}=0$ and $V=0$ baryon wave functions are
$$
-2\tilde\Sigma^{0}_{u\leftrightarrow s}(ds,u)=
|\Sigma^{0}(ud,s)>-\sqrt{3}|\Lambda>,
$$
$$
2|\tilde\Lambda_{u\leftrightarrow s}>=\sqrt{3}|\Sigma^{0}(ud,s)>+|\Lambda>.
$$
It is easy to show that relations given by 
Eqs.(\ref{sigma},\ref{lambda})
immeaditely follow.

\section{Relation between QCD correlators
for $\Sigma^{0}$ and $\Lambda$ hyperons}

Now we demonstrate how similar considerations work for QCD sum rules
on the example of QCD Borel mass sum rules.

The starting point would be two-point Green's function
for hyperons $\Sigma^{0}$ and $\Lambda$ of the baryon octet:
\begin{equation}
\Pi^{\Sigma^{0},\Lambda}=i\int d^{4}x e^{ipx}
<0|T\{ {\eta^{\Sigma^{0},\Lambda}(x),\eta^{\Sigma^{0},\Lambda}(0)}\}|0>,
\label{two}
\end{equation}
where isovector (with $I_{3}=0$) and isocalar field operators 
could be chosen as \cite{Altug1}
\begin{eqnarray}
\eta^{\Sigma^{0}}=\frac{1}{\sqrt2}\epsilon_{abc}
[\left(u^{aT}Cs^{b}\right)\gamma_{5}d^{c}+
\left(d^{aT}Cs^{b}\right)\gamma_{5}u^{c}-
\nonumber\\
\left(u^{aT}C\gamma_{5}s^{b}\right)d^{c}-
\left(d^{aT}C\gamma_{5}s^{b}\right)u^{c}],
\nonumber\\
\eta^{\Lambda}=\frac{1}{\sqrt{6}}\epsilon_{abc}
[-2\left(u^{aT}Cd^{b}\right)\gamma_{5}s^{c}-
\left(u^{aT}Cs^{b}\right)\gamma_{5}d^{c}
\nonumber\\
+\left(d^{aT}Cs^{b}\right)\gamma_{5}u^{c}+
2\left(u^{aT}C\gamma_{5}d^{b}\right)s^{c}
\\
+\left(u^{aT}C\gamma_{5}s^{b}\right)d^{c}-
\left(d^{aT}C\gamma_{5}s^{b}\right)u^{c}],
\nonumber
\hspace{10mm}
\label{eta}
\end{eqnarray}
where $a,b,c$ are the color indices and $
u,d,s$ are quark wave functions, $C$ is charge conjugation matrix,

We show now that one can operate with $\Sigma$ hyperon
and obtain the results
for the $\Lambda$ hyperon. The reasoning would be valid also for 
charm and beaty $\Sigma$-like and  $\Lambda$-like baryons.

In order to arrive at the desired relations we write not
only isospin quantities but also  $U$-spin and  $V$-spin ones.

Let us introduce $U$-vector 
(with $U_{3}=0$) and $U$-scalar field operators just
formally changing $(d\leftrightarrow s)$ in the Eq.(\ref{eta}):
\begin{eqnarray}
\tilde\eta^{\Sigma^{0}(d\leftrightarrow s)}
=
\frac{1}{\sqrt2}\epsilon_{abc}
[\left(u^{aT}Cd^{b}\right)\gamma_{5}s^{c}+
\left(s^{aT}Cd^{b}\right)\gamma_{5}u^{c}
\nonumber\\
-\left(u^{aT}C\gamma_{5}d^{b}\right)s^{c}-
\left(s^{aT}C\gamma_{5}d^{b}\right)u^{c}],
\nonumber\\
\tilde\eta^{\Lambda(d\leftrightarrow s)}=
\frac{1}{\sqrt{6}}\epsilon_{abc}
[-2\left(u^{aT}Cs^{b}\right)\gamma_{5}d^{c}-\qquad
\nonumber\\
\left(u^{aT}Cd^{b}\right)\gamma_{5}s^{c}
+\left(s^{aT}Cd^{b}\right)\gamma_{5}u^{c}+
2\left(u^{aT}C\gamma_{5}s^{b}\right)d^{c}
\nonumber\\
\hspace{10mm}+\left(u^{aT}C\gamma_{5}s^{b}\right)d^{c}-
\left(s^{aT}C\gamma_{5}d^{b}\right)u^{c}],
\label{tildeta}
\end{eqnarray}
Similarly we introduce $V$-vector 
(with $V_{3}=0$) and $V$-scalar field operators just
changing $(u\leftrightarrow s)$ in the Eq.(\ref{eta}):
\begin{eqnarray}
\tilde\eta^{\Sigma^{0}(u\leftrightarrow s)}=
\frac{1}{\sqrt2}\epsilon_{abc}
[\left(s^{aT}Cu^{b}\right)\gamma_{5}d^{c}+
\left(d^{aT}Cu^{b}\right)\gamma_{5}s^{c}
\nonumber\\
-\left(s^{aT}C\gamma_{5}u^{b}\right)d^{c}-
\left(d^{aT}C\gamma_{5}u^{b}\right)s^{c}],
\nonumber\\
\tilde\eta^{\Lambda(u\leftrightarrow s)}=
\frac{1}{\sqrt{6}}\epsilon_{abc}
[-2\left(s^{aT}Cd^{b}\right)\gamma_{5}u^{c}-\qquad
\nonumber\\
\left(s^{aT}Cu^{b}\right)\gamma_{5}d^{c}
+\left(d^{aT}Cu^{b}\right)\gamma_{5}s^{c}+
2\left(s^{aT}C\gamma_{5}d^{b}\right)u^{c}
\nonumber\\
+\left(s^{aT}C\gamma_{5}u^{b}\right)d^{c}-
\left(d^{aT}C\gamma_{5}u^{b}\right)s^{c}],
\hspace{10mm}
\label{tildeta2}
\end{eqnarray}
Field operators of the Eq.(\ref{eta}) and Eq.(\ref{tildeta})
can be related through
\begin{eqnarray}
-2\tilde\eta^{\Lambda(d\leftrightarrow s)} &=&
\eta^{\Lambda}-\sqrt{3}\eta^{\Sigma^{0}},
\nonumber\\
- 2\tilde\eta^{\Sigma^{0}(d\leftrightarrow s)} &=&
\sqrt{3}\eta^{\Lambda}+\eta^{\Sigma^{0}},
\nonumber\\
2\tilde\eta^{\Lambda(u\leftrightarrow s)} &=&
\eta^{\Lambda}+\sqrt{3}\eta^{\Sigma^{0}},
\\
2\tilde\eta^{\Sigma^{0}(u\leftrightarrow s)} &=&
\sqrt{3}\eta^{\Lambda}-\eta^{\Sigma^{0}},
\nonumber
\label{etarel}
\end{eqnarray}

Upon using Eqs.(\ref{eta},\ref{etarel})
two-point Green functions of the Eq.(\ref{two}) for 
hyperons $\Sigma^{0}$ and $\Lambda$ of the baryon octet
can be related as
\begin{eqnarray}
2[\tilde\Pi^{\Sigma^{0}(d\leftrightarrow s)}+
\tilde\Pi^{\Sigma^{0}(u\leftrightarrow s)}]-
\Pi^{\Sigma^{0}} &=&3\Pi^{\Lambda},
\label{crossSi}\\
2[\tilde\Pi^{\Lambda(d\leftrightarrow s)}+
\tilde\Pi^{\Lambda(u\leftrightarrow s)}]-
\Pi^{\Lambda}&=&3\Pi^{\Sigma^{0}}.
\label{crossLa}
\end{eqnarray}
These are essentially nonlinear relations.
 
It is seen that starting calculations, e.g., from 
$\Sigma$-like quantities one arrives at the correspondong quantities
for $\Lambda$-like baryons and {\it vice versa}.

It should be noted that since the overall normalizations of the currents depend on the convention,
in Eqs. (\ref{crossSi}) and (\ref{crossLa}), there is an ambiguity in these relations in the ratio of the coefficients 
of the correlators obtained from  $\Sigma^0$ correlator and lambda correlator. This ambiguity
results in the freedom to multiply the LHS {\em or} the RHS on {\em only one of} the Eqs.
(\ref{crossSi}) and (\ref{crossLa}) by an arbitrary constant. Once this is done on one of the
relations, the coefficients in the other relation are fixed. In Eq. (\ref{eta}), the normalization
is chosen so that the obtained relations for the correlators resemble the relations obtained for the
magnetic moments in NRQM, Eq. (\ref{nrqmrelations})
\section{Intercrossed relations for the QCD Borel sum rules}

In order to see how it works, we prefered not to use the results of one of us 
with coautors in \cite{Altug1,Altug}, which also satisfy our relations, but
we have repeated calculations of the first of the QCD mass sum rules 
for the
$\Sigma^{0}$ hyperon following \cite{Hwang}.  conserving 
non-degenerated quantities for $u$ and $d$ quarks, namely
\begin{eqnarray}
\frac{M^{6}}{8}L^{-4/9}E_{2}+\frac{bM^{2}}{32}L^{-4/9}E_{0}+
\frac{a_{u}a_{d}}{6}L^{4/9}-
\nonumber\\
\frac{a_{u}a_{d}(m_{0(u)}^{2}+m_{0(d)}^{2})}{48M^{2}L^{2/27}}-
\frac{m_{s}a_{s}m_{0(s)}^{2}}{24L^{26/27}}-
\nonumber\\
\frac{M^{2}E_{0}}{4L^{4/9}}[a_{s}m_{s}-(a_{u}-a_{d})(m_{d}-m_{u})]-
\nonumber\\
\frac{1}{48}[3m_{u}a_{d}m_{0(d)}^{2}+3m_{d}a_{u}m_{0(u)}^{2}-
m_{u}a_{u}m_{0(u)}^{2}
\nonumber\\
-m_{d}a_{d}m_{0(d)}^{2}]L^{-26/27}=
\beta^{2}_{\Sigma^{0}}e^{-(M^{2}_{\Sigma^{0}}/M^{2})}+e.s.c.,
\hspace{10mm}
\label{hwangSi}
\end{eqnarray}
where \cite{Hwang1}
\begin{eqnarray}
a_{q}=-(2\pi)^{2}<\bar q q>,\quad
b=<g_{c}G^{2}>,
\nonumber\\
a_{q}m_{0(q)}^{2}=(2\pi)^{2}<g_{c}\bar q \sigma\cdot G q>,\quad q=u,d,s.
\nonumber\\
L=ln(M^{2}/\Lambda^{2}_{QCD})/ln(\mu^{2}/\Lambda^{2}_{QCD}),
\label{vev}
\end{eqnarray}
$$
E_{n}(x)=1-e^{-x}(1+x+...+x^{n}/n!),
$$
$$
\quad x=W_{B}^{2}/M^{2}, \quad 
B=\Sigma^{0},\Lambda.
$$
Borel residue for the $\Lambda$ hyperon is defined as
$$
<0|\eta^{\Lambda}(0)|\Lambda(p)>=\lambda_{\Lambda}u(p),
\quad \beta^{2}_{\Lambda}=(2\pi)^{4}\lambda_{\Lambda}^{2},
$$
and similarly for the $\Sigma$ hyperon, while e.s.c. means
'excited-state contributions'.

The renormalization scale $\mu$ is taken usually to be around
1 GeV while QCD scale parameter should be  around 100 MeV.
With $m_{0(u)}=m_{0(d)}$ in Eq.(\ref{hwangSi})
one returns to the expression given by Eq.(21)
in \cite{Hwang}.

Now changing $(d\leftrightarrow s)$ and $(u\leftrightarrow s)$ in the 
LHS $(\Sigma^{0})$ of the Eq.(\ref{hwangSi}) to obtain 
LHS $(\bar \Sigma^{0}(d\leftrightarrow s))$ and
LHS $(\bar \Sigma^{0}(u\leftrightarrow s))$, respectively,
and using Eq.(\ref{crossSi}) we obtain for the $\Lambda$-mass SR:
\begin{eqnarray}
\frac{M^{6}}{8L^{4/9}}E_{2}+
\frac{bM^{2}}{32L^{4/9}}E_{0}+
\frac{2a_{s}(a_{u}+a_{d})- a_{u}a_{d}}{18}L^{4/9}-
\nonumber\\
\frac{L^{-2/27}}{144M^{2}}[2(a_{u}+a_{d})a_{s}m_{0(s)}^{2}+
2(m_{0(u)}^{2}a_{u}+m_{0(d)}^{2}a_{d})-
\nonumber\\
a_{u}a_{d}(m_{0(u)}^{2}+m_{0(d)}^{2})]-
\frac{M^{2}E_{0}m_{s}}{12L^{4/9}}[3a_{s}-2(a_{u}+a_{d})]
\nonumber\\
-\frac{M^{2}}{12L^{4/9}}E_{0}
[3(m_{u}a_{u}+m_{d}a_{d})+m_{d}a_{u}+m_{u}a_{d}-
\nonumber\\
2(m_{u}+m_{d})a_{s}]-
\frac{1}{48}[2m_{s}(a_{u}m_{0(u)}^{2}+a_{d}m_{0(d)}^{2})-
\nonumber\\
(a_{u}m_{0(u)}^{2}-a_{d}m_{0(d)}^{2})(a_{u}-a_{d})]L^{-26/27}-
\\
\frac{1}{24}(m_{u}+m_{d}-m_{s})a_{s}m_{0(s)}^{2}L^{-26/27}=
\beta^{2}_{\Lambda}e^{-(M^{2}_{\Lambda}/M^{2})}+e.s.c.
\nonumber
\label{hwangLa}
\end{eqnarray}
With $m_{0(u)}=m_{0(d)}=m_{0}$ in Eqs.(\ref{hwangLa})
one returns to the expressions given by Eq.(23)
in \cite{Hwang}.

If also  $a_{0(u)}=a_{0(d)}=a$, $m_{0}^{2}=m_{0(s)}^{2}$
and $m_{u}=m_{d}=0$ , one returns to mass
sum rules of  \cite{Ioffe} in the form given by Eq.(3) in \cite{Pasu3}
upon neglecting factors $L^{-2/27}$ and $L^{-26/27}$ 
in two terms at the LHS:
\begin{eqnarray}
\frac{M^{6}}{8L^{4/9}}+\frac{bM^{2}}{32}L^{-4/9}+\frac{a^{2}}{6}L^{4/9}-
\frac{a^{2}m_{0}^{2}}{24M^{2}L^{2/27}}-\\
\frac{a_{s}m_{s}M^{2}}{4L^{4/9}}-\frac{m_{s}a_{s}m_{0}^{2}}{24L^{26/27}}=
\beta^{2}_{\Sigma^{0}}e^{-(M^{2}_{\Sigma^{0}}/M^{2})}+e.s.c.,
\nonumber
\end{eqnarray}
\begin{eqnarray}
\frac{M^{6}}{8L^{4/9}}+\frac{bM^{2}}{32L^{4/9}}-
\frac{a_{s}^{2}m_{0}^{2}}{24M^{2}L^{2/27}}+
\\
\frac{a_{s}m_{s}M^{2}}{12L^{4/9}}-\frac{m_{s}a_{s}m_{0}^{2}}{24L^{26/27}}=
\beta^{2}_{\Lambda}e^{-(M^{2}_{\Lambda}/M^{2})}+e.s.c..
\nonumber
\end{eqnarray}
Now we show the validity of the second intercrossed 
relation Eq.(\ref{crossLa}).

We start now from the QCD Borel mass sum rule for the $\Lambda$
given by the Eq.(24) in \cite{Hwang}:
\begin{eqnarray}
\frac{M^{4}}{12}(2a_{u}+2a_{d}-a_{s})E_{1}-\frac{b}{16}(2a_{u}+2a_{d}-a_{s})+
\nonumber\\
\frac{\alpha_{s}}{\pi}\frac{L^{-1/9}}{243M^{2}}[108a_{u}a_{d}a_{s}+
a_{s}(a_{u}^{2}+a_{d}^{2})
\nonumber\\
-2(a_{u}a_{d}+a_{s}^{2})(a_{u}+a_{d})]
\nonumber\\
(\frac{M^{6}E_{2}}{12L^{8/9}}-
\frac{bM^{2}E_{0}}{96L^{8/9}})(2m_{u}+2m_{d}-m_{s})+
\nonumber\\
\frac{1}{36}[12m_{s}a_{u}a_{d}-2m_{s}a_{s}(a_{u}+a_{d})]+
\\
\frac{1}{36}[12a_{s}(m_{u}a_{d}+m_{d}a_{u})+
a_{s}(m_{u}a_{u}+m_{d}a_{d})
\nonumber\\
-2(m_{u}+m_{d})a_{u}a_{d}]=
\beta_{\Lambda}^{2}M_{\Lambda}e^{-(M^{2}_{\Lambda}/M^{2})}
+e.s.c.
\nonumber
\label{lambda2}
\end{eqnarray}
Performing changes $s\leftrightarrow d$ ($u\leftrightarrow d$)
we arrive at the corresponding Borel sum rules for
$\tilde\Lambda_{d\leftrightarrow s}$ 
($\tilde\Lambda_{u\leftrightarrow s}$).

Putting these expressions into Eq.(\ref{crossLa}) it is straightforward
to obtain
\begin{eqnarray}
\frac{a_{s}M^{4}}{4}E_{1}-\frac{a_{s}b}{72}+
\frac{\alpha_{s}}{\pi}\frac{L^{-1/9}}{81M^{2}}[-(a_{u}^{2}+a_{d}^{2})+
\nonumber\\
36a_{u}a_{d}]a_{s}+\frac{M^{6}}{4L^{8/9}}m_{s}E_{2}-
\frac{bM^{2}}{32L^{8/9}}m_{s}E_{0}+
\\
\frac{1}{12}a_{s}(4m_{u}a_{d}+4m_{d}a_{u}-m_{u}a_{u}-m_{d}a_{d})
\nonumber\\
+\frac{1}{3}m_{s}a_{u}a_{d}=
\beta^{2}_{\Sigma^{0}}M_{\Sigma^{0}}e^{-(M^{2}_{\Sigma^{0}}/M^{2})}
e.s.c.,
\nonumber
\label{crossSi2}
\end{eqnarray}
which is just the relation given by the Eq.(22) in \cite{Hwang}.
\section{Conclusion}

We have shown 
that starting from the QDC Borel mass sum rules for 
the $\Sigma$ hyperon
it is straightforward to obtain the corresponding
quantities for the $\Lambda$ hyperon and {\it vice versa}
upon using intercrossed relations of the type given by
Eqs.(\ref{sigma},\ref{lambda}) and Eqs.(\ref{crossSi},\ref{crossLa}).

More generally
these relations can be used not only to obtain properties
of the the $\Sigma$-like baryons
from those of $\Lambda$-like ones and {\it vice versa}
but also to check reciprocally many-terms relations for
the $\Sigma$-like and $\Lambda$-like baryons.
\section*{Acknowledgments}
We are grateful to T.Aliev, V.Dubovik, F.Hussain, N.Paver, G.Thompson
for discussions. We are also grateful to B. L. Ioffe for illuminating discussions and comments on
the final version of our work.
One of us (V.Z.) is grateful to Prof.S.Randjbar-Daemi
for the hospitality extended to him at HE section of ICTP 
(Trieste, Italy).

\end{document}